# Frequency comb-based microwave transfer over fiber with 7×10$^{-19}$ instability using fiber-loop optical-microwave phase detectors

Kwangyun Jung,[1] Junho Shin,[1] Jinho Kang,[1] Stephan Hunziker,[2] Chang-Ki Min,[3] and Jungwon Kim[1,*]

[1]Korea Advanced Institute of Science and Technology (KAIST), Daejeon 305-701, South Korea
[2]Paul Scherrer Institute (PSI), Villigen 5232, Switzerland
[3]Pohang Accelerator Laboratory (PAL), Pohang 790-834, South Korea
*Corresponding author: jungwon.kim@kaist.ac.kr

We demonstrate a remote microwave/radio-frequency (RF) transfer technique based on the stabilization of a fiber link using a fiber-loop optical-microwave phase detector (FLOM-PD). This method compensates for the excess phase fluctuations introduced in fiber transfer by direct phase comparison between the optical pulse train reflected from the remote site and the local microwave/RF signal using the FLOM-PD. This enables sub-fs resolution and long-term stable link stabilization while having wide timing detection range and less demand in fiber dispersion compensation. The demonstrated relative frequency instability between 2.856-GHz RF oscillators separated by a 2.3-km fiber link is 7.6×10$^{-18}$ and 6.5×10$^{-19}$ at 1000 s and 82500 s averaging time, respectively.

High-precision transfer and remote synchronization of microwave and radio-frequency (RF) signals over long distance is important for many important scientific and technical applications such as dissemination of time and frequency standards, navigation and timekeeping, synchronization of large-scale scientific facilities (such as accelerator-based light sources and long-baseline radio astronomy), remote and distributed sensing, and telecommunication networks, just to name a few. In the last decade, there have been remarkable progresses in the precise transfer and synchronization of microwave/RF signals over optical fiber links. Two representative methods are (i) transfer of microwaves from modulated continuous-wave (CW) lasers and (ii) transfer of optical pulse trains from femtosecond mode-locked lasers. Modulated CW transfer had recently showed a relative frequency instability of ~10$^{-17}$ (2×10$^{-17}$ [1] and 7×10$^{-18}$ [2]) at 1000 s averaging time and ultimately reached ~4-5×10$^{-19}$ (5×10$^{-19}$ [1] and 4×10$^{-19}$ [2]) at 70000 s averaging time. Direct transfer of frequency combs can also carry microwave/RF signal encoded in the pulse repetition rates, which recently resulted in 4-7×10$^{-17}$ (7×10$^{-17}$ [3], 4×10$^{-17}$ [4], 6×10$^{-17}$ [5]) at 1000 s averaging time and reached 6.6×10$^{-18}$ [5] at 16000 s averaging time.

In both optical transfer methods, the most important task is to compensate for the excess phase fluctuations introduced in the long fiber links. Although direct transfer of frequency combs has an advantage that it can carry both ultrashort optical pulses and microwave/RF signals in a form of optical pulse trains, the limited timing/phase detection resolution and large temperature-dependent timing/phase drift in the optical-to-microwave/RF conversion by fast photodetection has limited its achievable timing and phase stability. To overcome this limitation, one can use an ultra-sensitive balanced optical cross-correlation (BOC) method as shown in [6] and [7], which has enabled (sub-)femtosecond-level fiber link stabilization. However, the use of BOC can be sometimes technically difficult, especially when the fiber links become longer, because it is alignment-sensitive, has a limited timing detection range (on the order of a few ps at most), and requires very precise link dispersion compensation as well as link length matching (which is, an exact multiple of pulse-to-pulse spacing). These limitations of BOC are due to the nature of nonlinear optic process (e.g., type-II phase-matched sum-frequency generation) where the timing detection is performed by cross-correlation between femtosecond optical pulses.

Therefore it would be highly desirable if one can compensate for the phase fluctuation in the transfer of frequency combs (optical pulse trains) via fiber links using a method with sub-fs resolution and long-term stability, which is better than the direct photodetection method, and at the same time with a larger timing detection range and less sensitivity to pulsewidth and fiber dispersion than the BOC method. For this purpose, our recently demonstrated fiber-loop optical-microwave phase detector (FLOM-PD) [8,9] can be used: at 10-GHz carrier frequency, the FLOM-PD could directly detect the relative phase between the optical pulse train and the microwave signal with both sub-fs resolution (0.6-fs resolution in 10-MHz bandwidth [9]) and sub-fs long-term phase stability (0.8-fs accumulated drift for 2 hours [8]). In this Letter, we demonstrate that the use of FLOM-PD for 2.3-km fiber link stabilization can result in frequency comb-based remote microwave (2.856 GHz in this work) transfer with 6.5×10$^{-19}$ level instability (at 82500-s averaging time) and 4.8-fs rms drift integrated over 8 hours. We also test the FLOM-PD-based link stabilization method with different fiber length, dispersion, and installation conditions, and show that ~10$^{-18}$ or lower instability can be consistently obtained. We believe that currently achievable stability in the comb-based microwave/RF transfer was mostly limited by the polarization mode dispersion (PMD) and polarization state drifts in the used fiber links when the fiber link temperature changes.

Figure 1 shows the experimental setup for demonstrating the microwave/RF transfer stabilized by the FLOM-PD. The brief operation principle is following. The optical pulse train from a mode-locked laser at the local site is transferred to the remote site via a fiber link. For the link stabilization, a microwave signal tightly synchronized with the mode-locked laser at the local site (using the FLOM-PD #1 in Fig. 1) is used. The excess phase fluctuation added in the fiber link transfer is detected by the FLOM-PD (FLOM-PD #2 in Fig. 1), which directly compares the microwave phase at the local site with the optical pulse train reflected back from the remote site by a partial reflector. By using a microwave signal for the phase error detection, one can extend the timing detection and locking acquisition range to tens – hundreds ps (depending on the used microwave frequency), in contrast to the BOC case of <1 ps range. The detection sensitivity and internal drift of FLOM-PD are still in the sub-fs level [8,9], which enables fs-level link stabilization. The detected phase error is fed back to the stage (piezoelectric transducer (PZT)-based and/or motorized) to counteract the excess phase noise introduced to the fiber link. At the remote site, microwave can be extracted from the delivered optical pulse train using the third FLOM-PD (FLOM-PD #3 in Fig. 1) and a high-quality voltage-controlled oscillator (VCO). As a result, the microwave signals at the local and remote sites are effectively synchronized via a stabilized fiber link.

More detailed information on the experimental setup is following. A low phase noise RF signal generator (Agilent N5181B) at 2.856 GHz is used as the master oscillator. In this work, 2.856-GHz frequency is chosen for the demonstration because it can be useful for driving multiple remotely located S-band accelerating cavities used for many free-electron lasers. In principle, any microwave/RF frequency components, which are the multiple of the laser repetition rate, can be used. A 238-MHz repetition rate, 120-fs pulsewidth soliton mode-locked Er-glass laser (Onefive GmbH, Origami-15) is used as the frequency comb source, which is locked to the master RF generator using the FLOM-PD #1. Here, 16 mW of optical power and +10 dBm of RF power are applied to the FLOM-PD #1. The mode-locked laser is synchronized to the RF generator using a PZT inside the laser cavity with 6-kHz locking bandwidth. The locking bandwidth is set to 6 kHz because the equivalent phase noise of the mode-locked laser is higher than that of the RF generator below 6 kHz offset frequency. As a result, we can take advantage of the flywheel effect in both low phase noise regions of RF generator and mode-locked laser (i.e., <6 kHz offset frequency range of the RF generator and >6 kHz range of the mode-locked laser).

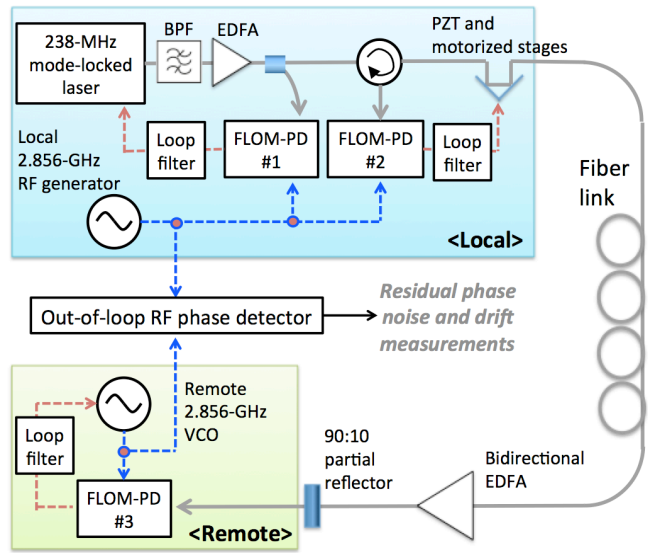

Fig. 1. Schematic of the FLOM-PD-stabilized fiber link experiment. For the fiber link, (i) 610-m long dispersion-compensated fiber link, (ii) 2.3-km long dispersion-compensated fiber link (installed in an accelerator building), (iii) 1-km long SMF-28 fiber link (without dispersion compensating fiber), and (iv) 8-m long SMF-28 fiber link are used. Note that a 2.3-nm bandpass filter (BPF) is used for (i) and (iii) only.

About 2 mW of optical power is applied to the fiber link. In this work, four different kinds of fiber links are used for the test: (i) 610-m long, dispersion compensated fiber link (450-m SMF-28 and 160-m dispersion compensating fiber (DCF, Thorlabs DCF38) in a spool), (ii) 2.3-km long, dispersion compensated fiber link (2.1-km SMF-28 and 200-m DCF (OFS LLWB); 2.2-km in a spool and 100-m installed in a cable duct in an accelerator building), (iii) 1-km long SMF-28 fiber link (in a spool) without dispersion compensation, and (iv) 8-m long SMF-28 fiber link (to assess the performance of FLOM-PDs only). Note that, for the 1-km link, we intentionally did not use DCF to show that FLOM-PD can be still used for link stabilization with even ~20-ps pulsewidth (after stretched by 2-km round-trip fiber length). To avoid excessive chirp in the 1-km fiber link, a 2.3-nm bandpass filter (BPF) at 1540 nm is used. Also note that for the 2.3-km long fiber link, part of the link (~100 m) is installed in the cable duct in a linear accelerator building of the Pohang Accelerator Laboratory (where the temperature change is measured to be ~3 K in 4 days) to show that this fiber link also works in an accelerator environment. At the end of fiber link, a bidirectional Er-doped fiber amplifier (EDFA) is used to compensate for splicing losses between the SMF-28 and the DCF. For the 2.3-km link, ~0.3 mW of delivered power is amplified to 8 mW by the link EDFA. Part of the pulse train (10 %) is reflected back by a partial reflector at the remote location, redirected by a circulator, and compared with the local 2.856-GHz signal using the FLOM-PD #2. The used optical and RF power are 2 mW and +10 dBm, respectively. The detected error signal, which is proportional to the excess phase noise introduced in the link transfer, is fed back to the PZT and motorized stages. We found that, depending on the used PZT and motor stages, the power and polarization state changes can be severe, which limits the link performance. Note that, for us, the retro-reflector mounted on the combination of the PZT stage with 120-μm range (PI, P-621.10L) and the motorized stage with 15-mm range (PI, M-111.12S) provided the best power and polarization state stability for the link.

At the remote site, the 7-mW delivered pulse train is synchronized with the +21 dBm, 2.856-GHz voltage controlled oscillator (VCO, INWAVE DRO-2856A) using the FLOM-PD #3 with ~100 kHz locking bandwidth. This completes the remote transfer of 2.856-GHz RF signal. In order to compare the phase stability between the local and remote RF sources, the local and remote setups are built closely each other and an out-of-loop RF phase detector (which is shown in [10]) is used for measuring the residual phase noise and drift. A special effort was given to the construction of this long-term stable RF phase detector, especially to enable the sub-10-fs resolution long-term RF phase measurement over many days.

Figure 2 shows the measured residual phase noise between the local and remote 2.856-GHz RF sources. Curve (a) shows the phase noise when a 2.3-km link is used. Curve (b) shows the phase noise when an 8-m link is used. Curve (c) is the phase noise of the free-running VCO. As the locking bandwidth is limited to ~100 kHz in the VCO, the achievable residual noise floor is limited to ~ -130 dBc/Hz. The integrated rms timing jitters with and without 2.3-km link are 21.5 fs (14.5 fs when excluding 60-Hz peak from power supply) and 8.7 fs, respectively. Note that the broad peak around 100 Hz in curve (a) is caused by the limited bandwidth of the PZT (with ~250 Hz resonance frequency) used for the link stabilization. As a result, insufficiently suppressed acoustic noise in the link is remained. Provided a faster PZT for link stabilization, this residual noise could have been further suppressed.

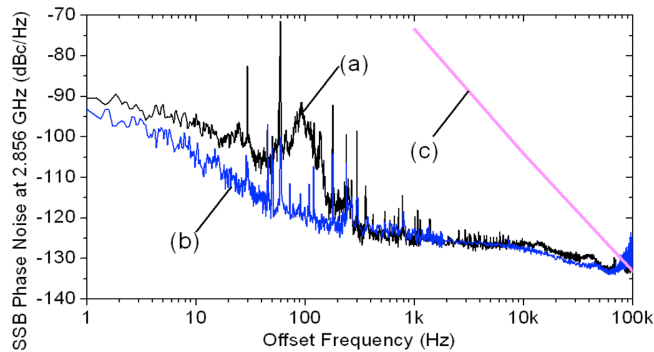

Fig. 2. Phase noise data. (a) Residual phase noise between the 2.856-GHz oscillators when a stabilized 2.3-km link is used. (b) Residual phase noise with an 8-m link. (c) Phase noise of the free-running VCO.

Figure 3 shows the collection of long-term residual phase drift measurements between the remote 2.856-GHz oscillators. Note that 1-Hz low-pass filtering is used at the output of the RF phase detector and the phase information is recorded with 1 sample/s rate. Figs. 3(a), 3(b), and 3(c) show the 2.3-km link, 1-km link (without DCF), and 8-m link results, respectively. For the 2.3-km link [Fig. 3(a)], the integrated rms timing drift is 36 fs over 92 hours, including the 8-hour region with 4.8 fs drift (indicated by a red-lined box in Fig. 3(a)). The PZT-plus-motor stage compensated for the drift of >50 ps over 4 days. For the 1-km link without dispersion compensation [Fig. 3(b)], 108 fs rms drift is measured over 42 hours (with 27 fs rms drift for the first 10 hours). Note that for the 1-km link, there was periodic temperature fluctuation with ~15-min period in the laboratory, which is also shown in the timing drift and motor displacements as well. The worse performance compared to Fig. 3(a) may be caused by the long reflected pulsewidth (>20 ps), which reduces the FLOM-PD phase detection sensitivity in the link stabilization. Also note that the amplitude-to-phase conversion suppression technique in FLOM-PD [11] had to be employed in the 1-km link case, as longer and more distorted optical pulses are more susceptible to the amplitude-to-phase conversion in FLOM-PDs. To assess the stabilization and synchronization performances of FLOM-PDs only, we also measured the phase drift of very short (8 m in this work) fiber link, which is shown in Fig. 3(c). It shows 5.3 fs rms drift integrated over 10 hours.

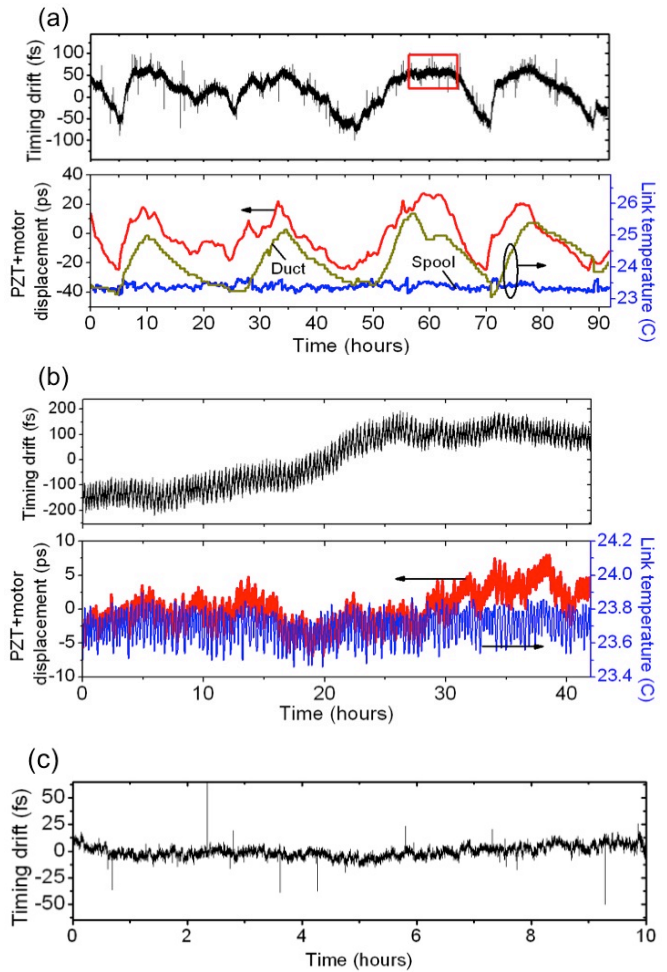

Fig. 3. Phase drift measurement results. (a) 2.3-km dispersion compensated fiber link (without BPF). (b) 1-km fiber link without DCF (with BPF). (c) 8-m fiber link (without BPF).

For the 2.3-km link, the residual phase drift could be maintained to ~5-fs rms stability up to 10 h. However, as shown in Fig. 3(a), over longer periods of time there are phase drifts of ~100 fs, that closely follow the link temperature change (and motor displacement), with a ~1-day period. Given that the loop gain by the proportional-integral (PI) servo should be sufficiently high at this offset frequency (tens µHz), we believe that this longer-term drift might be caused by the polarization state drifts and PMD in the fiber link, originated from fiber temperature (and birefringence) changes. Similar drift has been observed in the previous measurements of 300-m long fiber links using the BOC, as shown in [12]. In our case, the fact that the FLOM-PD is based on PM-fibers (and sensitive to the polarization state changes) seem to make the polarization-drift effects more significant.

Figure 4 shows the calculated relative frequency instability based on the phase drift measurements, in terms of overlapping Allan deviation. Curves (a), (b), (c) and (d) indicate the 2.3-km link, 610-m link, 1-km link, and 8-m link data, respectively. The resulting frequency instabilities at 1000 s for dispersion-compensated 610-m and 2.3-km links are $8.5 \times 10^{-18}$ and $7.6 \times 10^{-18}$, respectively. Ultimately, the frequency instability reaches $6.5 \times 10^{-19}$ at 82500 s averaging time for the 2.3-km fiber link. These instability levels are an order of magnitude improved from the previous comb-based microwave transfer results. Note that the instability could reach $1.7 \times 10^{-18}$ even without dispersion compensation for the 1-km fiber link.

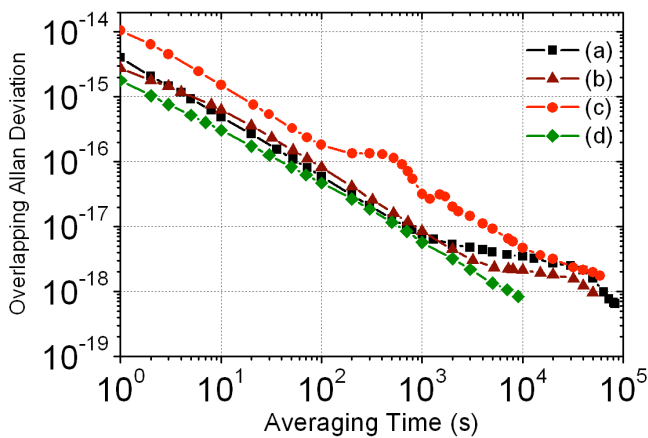

Fig. 4. Relative frequency instability between remote 2.856-GHz microwave oscillators in terms of overlapping Allan deviation. (a) dispersion-compensated 2.3-km fiber (without BPF). (b) dispersion-compensated 610-m fiber (with BPF). (c) 1-km fiber without dispersion compensation (with BPF). (d) 8-m long fiber (without BPF) for reference.

In summary, we demonstrate that the use of all-fiber FLOM-PDs for fiber link stabilization enables frequency comb-based remote microwave/RF transfer over 2.3-km link with $6.5 \times 10^{-19}$ level instability (at 82500-s averaging time) and 4.8-fs drift integrated over 8 hours. As the currently demonstrated phase drift and instability might be mostly limited by the polarization drifts and PMD caused by fiber temperature changes, and not by the FLOM-PD performance itself, the mitigation of polarization issues may further improve the transfer performance. As a future work, we will first investigate whether the use of a Faraday rotating mirror at the link output [6] can mitigate some of the polarization issues in the link. The use of higher microwave frequency (e.g., 8 GHz in [3,4]) may also improve the timing drift (provided the same phase drift) and relative instability as well. For a few-km link (suitable for transfer in large-scale facilities such as FELs), a PM-fiber link as shown in [7] can be employed with FLOM-PD stabilization scheme. Note that the all-PM-fiber nature and less demand for dispersion compensation of FLOM-PD will fit well with the PM-fiber links. For much longer-range (over tens to hundreds km) transfer of combs with <$10^{-18}$-level microwave instability, we have an interest in exploring polarization control at the fiber link output [13] and developing a new schematic with polarization scrambling and a polarization-insensitive optical-microwave phase detector. We also note that the FLOM-PD-based timing stabilization might be particularly useful for free-space timing links [14,15], where the amount of timing/phase fluctuations is larger while the chromatic dispersion, polarization drift and PMD are smaller than the fiber links.

This research was supported in part by the National Research Foundation (Grant 2012R1A2A201005544) and the PAL-XFEL Project of South Korea.